# Fully Convolutional Slice-to-Volume Reconstruction for Single-Stack MRI

Sean I. Young*
Harvard Medical School
siyoung@mit.edu

Yaël Balbastre*
Harvard Medical School
ybalbastre@mgh.harvard.edu

Bruce Fischl
Harvard Medical School
fischl@nmr.mgh.harvard.edu

Polina Golland
MIT
polina@csail.mit.edu

Juan Eugenio Iglesias
Harvard Medical School
jei@mit.edu

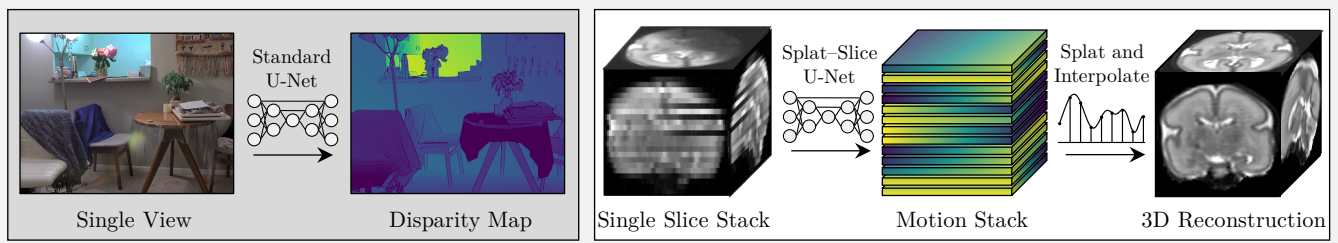

(a) Monocular Depth (Disparity) Estimation [38–41]    (b) Single-Stack Slice-to-Volume Reconstruction (Proposed)

**Figure 1: Method overview.** Inspired by monocular disparity estimation (a), we tackle slice-to-volume reconstruction (b) by predicting slice motion from a single stack of 2D slices using a fully convolutional network model. The intensities of the 2D slices are splatted using the slice motion predicted by our network and the splatting is then interpolated to produce an artifact-free 3D MR reconstruction.

## Abstract

*In magnetic resonance imaging (MRI), slice-to-volume reconstruction (SVR) refers to computational reconstruction of an unknown 3D magnetic resonance volume from stacks of 2D slices corrupted by motion. While promising, current SVR methods require multiple slice stacks for accurate 3D reconstruction, leading to long scans and limiting their use in time-sensitive applications such as fetal fMRI. Here, we propose a SVR method that overcomes the shortcomings of previous work and produces state-of-the-art reconstructions in the presence of extreme inter-slice motion. Inspired by the recent success of single-view depth estimation methods, we formulate SVR as a single-stack motion estimation task and train a fully convolutional network to predict a motion stack for a given slice stack, producing a 3D reconstruction as a byproduct of the predicted motion. Extensive experiments on the SVR of adult and fetal brains demonstrate that our fully convolutional method is twice as accurate as previous SVR methods. Our code is available at* `github.com/seannz/svr`.

## 1. Introduction

For non-invasive imaging of the brain where the subject may potentially exhibit severe uncontrollable motion—such as in the case of the fetal population—two-dimensional (2D) magnetic resonance imaging (MRI) techniques are typically used to acquire a stack of 2D brain slices and assemble them into a 3D volume, a computational procedure often referred to as slice-to-volume reconstruction (SVR). Unlike 3D MRI acquisitions where the entire $k$-space (Fourier) coefficients can become corrupted even from a single motion event, 2D techniques can restrict this corruption to within a slice while leaving the $k$-space data for the other slices intact. The SVR procedure can then be applied on the non-corrupted slices to enable artifact-free imaging of the brain and facilitate other downstream tasks, such as brain morphometry [1–3], whole brain segmentation [4–6] and atlas creation [7–9].

A large number of classical SVR techniques [10–18] are descended from the work of Rousseau et al. [19], where the SVR task is posed as an optimization problem to recover an unknown 3D volume from observed 2D slices. Such inverse methods do not have the capability to learn the latent space of reconstructions, and as a result, typically require multiple orthogonal slice stacks to cross-regularize the estimation of the motion of each stack, adding a computational burden to an already iterative numerical method. More recent methods [20–25] overcome the burden of numerical optimization by directly predicting the position parameters of 2D slices with neural network models. However, these approaches process slices independently of each other and cannot constrain the reconstruction in 3D space [20–22] or require multiple slice stacks for accurate reconstruction [25], precluding their use in time-sensitive applications such as fetal fMRI [26, 27].

In the closely related tasks of pairwise image registration [28–32], motion [33–39] and depth estimation [40–43], fully convolutional network (FCN) architectures have been used to enable accurate non-iterative prediction of an underlying deformation, motion and disparity from a pair of images. By formulating the SVR task as a supervised motion estimation problem, we can reduce the task at hand to a problem that is more readily tackled using a convolutional framework. This convolutional approach to SVR bears a certain resemblance to monocular depth estimation, where disparity is predicted from a single view [42]. This is illustrated in Figure 1.

---

*These authors contributed equally to this work.

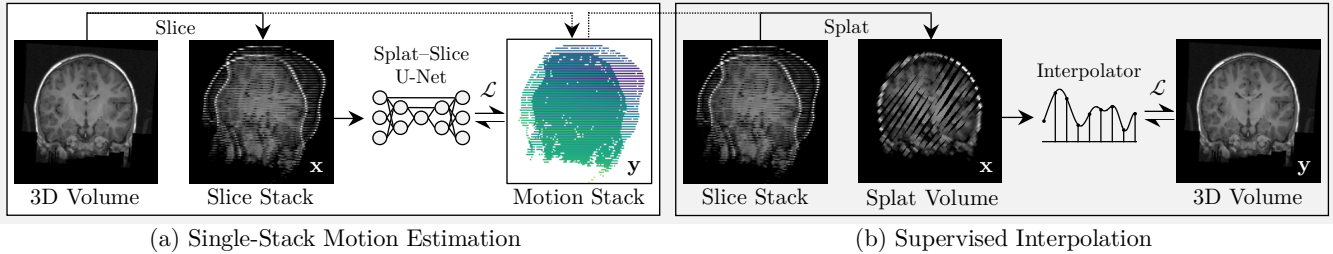

Figure 2: **Supervised learning.** We train our splat–slice U-Net (a) on paired (slice stack, motion stack) examples, which are generated by slicing 3D volumes with randomized slice-wise motion (motion stack). We additionally train a fully convolutional interpolation network (b) on paired (splat volume, 3D volume) examples, where splat volumes are generated by splatting slice stacks with ground truth motion.

In this work, we propose a fully convolutional method for SVR with application to single-stack MRI. We pose the SVR task as a supervised image registration problem and train a convolutional network on paired (slice stack, motion stack) data to predict slice motion directly, producing a 3D volume reconstruction as a result (Figure 2). Our approach requires only a single slice stack for reconstruction, is many orders of magnitude faster than optimization approaches [10–18], and reduces error in the predicted motion by $2\times$ compared to the recent transformer-based model [25]. Our contributions are:

- **Splat–Slice U-Net Model.** We interpret the iterations of classical SVR as a series of splat and slice operations to design a 3D FCN model for fast slice motion prediction.

- **Supervised Learning.** We formulate a parameter-free loss function to facilitate training with true slice motion for slice motion prediction with sub-voxel accuracy.

- **Artifact-Free Reconstruction.** We obtain artifact-free 3D reconstructions by splatting slice data and imputing missing intensities via supervised interpolation.

## 2. Related Work

In fetal imaging, slice-to-volume reconstruction (SVR) is an important inverse problem with reconstruction methods ranging from traditional optimization-based ones [10–19] to recent learning-based ones [20–25], which we now review.

### 2.1. Optimization Approaches

Optimization-based SVR [10–18] traces its way back to the seminal work of Rousseau et al. [19], who predicted the underlying 3D MRI volume by alternatingly estimating the slice motion and volume from three orthogonal slice stacks and super-resolving the final predicted volume. The method of Jiang et al. [10] is closely related to [19] but adopts a grid search to estimate the slice motion parameters to overcome the non-convexities of the optimization problem. Gholipour et al. [11] and Kuklisova-Murgasova et al. [13] attribute the cause of poor reconstructions to inaccurate slice alignments and propose reconstruction strategies to combat misaligned slices post-hoc. Uus et al. [18] extend [13] to accommodate deformable motion models as would be required, e.g., in the reconstruction of the placenta. While the focus of our work is on brain reconstruction, our supervised learning approach similarly accommodates different motion models (rigid and deformable) by encoding regularity of motion directly into the training motion fields without additional regularization.

Several SVR approaches extend [19] by reparametrizing slice motion for a more favorable computational complexity or flexibility of the slice motion. Kim et al. [12] propose to lower the cost of motion estimation by aligning 1D lines of intersection between 2D slices. Alansary et al. [15] propose to handle deformable slice motion by partitioning slices into square patches and estimating a separate set of rigid motion parameters for each patch. However, this strategy produces a patch-wise linear approximation of the underlying motion and leads to block artifacts in the reconstructed volume. In our work, we represent slice motion as dense motion fields to enable straight-forward artifact-free 3D reconstruction.

In optimization-based SVR methods, good initialization of the 3D volume and the slice motion can help mitigate the deleterious effects of local minima in the objective function on the reconstruction. Kainz et al. [14] propose techniques to automatically select a reference slice stack to improve the initialization of the 3D volume. Tourbier et al. [16] initialize the 3D volume with an age-matched template image, which acts as prior knowledge in the reconstruction framework. By contrast, we task our fully convolutional network model with robustly determining slice motion in the presence of image non-convexities, allowing it to dispense with templates and produce subject- and pathology-specific reconstructions.

### 2.2. Learning-Based Approaches

Currently, the majority of machine-learning-based SVR methods [20–25] predict the position parameters of slices (9

numbers) within the underlying MRI volume by using a 2D convolutional neural network (CNN). The predicted position parameters are then used to warp the slices and recover the underlying MRI volume. The CNN model of Hou et al. [20] consists of the convolutional layers of VGG-16 pre-trained on ImageNet [44] and a dense connected head to predict the anchor points (AP) of the individual slices. Yeung et al. [24] extend [20] and predict the position parameters of all slices jointly, introducing an inter-slice attention layer between the extracted features and the output to conditionally predict the motion for each slice given the features of other slices.

More recently, a non-convolutional model for SVR was proposed by Xu et al. [25] in an unsupervised setting using a vision transformer model [45–47]. Although this produces promising results, it scales quadratically with the number of slices due to inter-slice attention computations and requires hyperparameter tuning to balance different terms in the loss function. By posing the SVR task as a supervised registration problem within a convolutional framework, we significantly lessen the complexity of modeling pairwise slice interaction through convolutional weight sharing and obviate the need for hyperparameter optimization in the loss function.

The problem of regressing slice position (or even motion) against intensity data is highly non-convex due to the image landscape [19]. While there have been efforts to overcome image non-convexities in learning-based image registration [31, 32], these concerns are not addressed in learning-based pose prediction works [24, 25] and are exacerbated further by separate processing of the individual slices [20]. Posing SVR as image registration allows us to appeal to e.g. [31, 32] to address the fundamentally non-convex nature of SVR.

### 2.3. Pre- and Post-Processing

SVR techniques typically involve slice pre-processing as a first step to reject slices exhibiting considerable intra-slice motion and segment relevant, e.g., the brain, structures from the scan. Anquez et al. [48] propose a method to skull-strip fetal volumes, which can be used also to mask out non-brain structures in MR slices for small inter-slice motion. Tourbier et al. [16] propose to localize brain regions in MR slices by block-matching an age-matched template to the slices then registering the matched regions back to the template. Kainz et al. [49] localize brain regions based on rotation-invariant descriptors for labeling brain. Keraudren et al. [50, 51] use a dense SIFT and random forest classification for voxel-level segmentation. Some shortcomings of these methods include high computational complexity [16, 48] as well as the need to handcraft features [50, 51] for segmentation.

Learning-based approaches to fetal segmentation include the methods of Rajchl et al. [52] and Salehi et al. [53], both of which use FCN models [54] to predict foreground masks directly on slices. For post-processing of the reconstructed 3D volume, Xu et al. [55] propose neural volume rendering similar to NeRF [56] to super-resolve reconstructions at an arbitrary resolution. In this work, we additionally train an interpolation network to postprocess the raw reconstructions and interpolate over regions of missing intensities that may appear in the single-stack reconstruction case, producing an artifact-free reconstruction for other downstream tasks.

## 3. Mathematical Preliminaries

### 3.1. Slice Acquisition Model

In high-resolution 3D MRI of subjects exhibiting severe uncontrollable motion, fast slice acquisition sequences such as single shot fast spin echo are used to "freeze" the motion in-plane, substantially mitigating motion artifacts compared to multi-shot methods [57]. SVR procedures aim to align an acquired stack of slices thereby removing the remaining (in-plane and through-plane) motion across slices.

Let us denote the coordinates inside the MRI scanner by $(x, y, z)$ and assume a stack $f$ of 2D slices is acquired along the $z$ axis with spacing $s$ from an underlying volume $v$. The model for the acquisition of the $k$th MR slice is [11, 19]:

$$f_k(x, y) = \varepsilon + \int_{-\infty}^{\infty} (v \circ u_k)(x, y, z) h(sk - z) \, \mathrm{d}z \quad (1)$$

for $k = 1, \ldots, K$, where $u_k : (x, y, z) \mapsto (x, y, z)'$ denotes an unobserved slice-varying rigid spatial subject motion, $h$ denotes the point spread function of slice acquisition, and $\varepsilon$ captures all MR-induced errors such as noise, bias, etc.

To discretize model (1), suppose $v$ is sampled on a cubic lattice of $N^3$ points and $f_k$ on a 2D lattice of $N^2$ points. We can write the discretized model (up to random noise) as

$$\mathbf{f} = [\mathbf{f}_1; \ldots; \mathbf{f}_K] \in \mathbb{R}^{N^2 K}, \quad \mathbf{f}_k = \mathbf{H}_k \mathbf{U}_k \mathbf{v} \in \mathbb{R}^{N^2}, \quad (2)$$

in which $\mathbf{U}_k$ and $\mathbf{H}_k$ represent discretizations of $(\circ \, u_k)$ and $h(sk - \cdot)$, respectively. In the small motion regime, we can further appeal to Taylor's approximation and define

$$\mathbf{U}_k \mathbf{v} = \mathbf{v} + [\mathbf{diag}(\mathbf{v}_x), \mathbf{diag}(\mathbf{v}_y), \mathbf{diag}(\mathbf{v}_z)] \mathbf{S} \mathbf{u}_k, \quad (3)$$

in which $\mathbf{S} \mathbf{u}_k$ represents the expansion of the coefficients of motion $\mathbf{u}_k$ in a basis $\mathbf{S}$ and $\mathbf{v}_{\{x,y,z\}}$ are the respective spatial derivatives of the 3D volume. We accommodate deformable motion models by setting $\mathbf{S}$ to the identity matrix, in which case, $\mathbf{u}_k$ denotes the vectorization of a 3D motion field.

### 3.2. Classical SVR Methods

For brevity, let us write $\mathbf{U} = [\mathbf{H}_1 \mathbf{U}_1; \ldots; \mathbf{H}_K \mathbf{U}_K]$ and the associated motion parameters as $\mathbf{u} = [\mathbf{u}_1; \ldots; \mathbf{u}_K]$. We can write the SVR problem with three orthogonal stacks as

$$\begin{aligned}
\text{minimize} \quad & D(\mathbf{u}^{\{1,2,3\}}, \mathbf{v}^{\{1,2,3\}}) = \|\mathbf{U}^1 \mathbf{v}^1 - \mathbf{f}^1\|_2^2 \\
& + \|\mathbf{U}^2 \mathbf{v}^2 - \mathbf{f}^2\|_2^2 + \|\mathbf{U}^3 \mathbf{v}^3 - \mathbf{f}^3\|_2^2 \quad (4) \\
\text{subject to} \quad & \mathbf{v}^1 - \mathbf{v}^2 = \mathbf{v}^2 - \mathbf{v}^3 = \mathbf{v}^3 - \mathbf{v}^1 = \mathbf{0}
\end{aligned}$$

in which $\mathbf{u}^{\{1,2,3\}}$ denote the parameters of motion in each of the stacks. Contrast-invariant loss functions such as mutual information loss [58] are also popularly used in place of the quadratic one shown here for simplicity. Here, we introduce optimization variables $\mathbf{v}^{\{1,2,3\}}$ with equality constraints as opposed to a single variable $\mathbf{v}$ to facilitate solving (4) using alternating optimization strategies similarly to [17, 19].

A simple alternating optimization strategy which does not introduce dual variables is to relax the equality constraints to the quadratic penalty

$$R(\mathbf{v}^{\{1,2,3\}}) = \|[\mathbf{v}^1, \mathbf{v}^2, \mathbf{v}^3] - [\mathbf{v}^2, \mathbf{v}^3, \mathbf{v}^1]\|_2^2 \quad (5)$$

and optimize the relaxation

$$J(\cdot) = D(\mathbf{u}^{\{1,2,3\}}, \mathbf{v}^{\{1,2,3\}}) + \lambda R(\mathbf{v}^{\{1,2,3\}}), \quad (6)$$

in which $\lambda$ is a parameter controlling the relative weights of the data fidelity term $D$ and the coupling term $R$. Objective (6) in fact corresponds to the objective implicitly optimized by Rousseau et al. [19] when the mutual information loss is used in the data fidelity term $D$. Optimizing (6) using block coordinate descent yields updates for $\mathbf{v}^1, \mathbf{u}^1, \mathbf{v}^2, \mathbf{u}^2, \mathbf{v}^3$ and $\mathbf{u}^3$, where the 3D reconstructions ($v$-update) and the motion parameters ($u$-update) of each stack are updated in turn with the other two reconstructions fixed at each turn.

In the $n$th iteration of the block coordinate descent, the $v$-update step for a given stack amounts to computing

$$\mathbf{v}^n \leftarrow (\mathbf{U}^{n*}\mathbf{U}^n + 2\lambda\mathbf{I})^{-1}(\mathbf{U}^{n*}\mathbf{f}^n + 2\lambda\overline{\mathbf{v}}^n), \quad (7)$$

in which $\mathbf{v}^n$ and $\mathbf{f}^n$ denote the reconstruction and 2D slices of the $n\%3$rd stack, respectively, and $\overline{\mathbf{v}}^n$ denotes the average reconstruction from the two stacks orthogonal to $\mathbf{v}^n$. We can interpret this update as a reconstruction of a new 3D volume from a weighted sum of the last average 3D reconstruction and warped slices, followed by Wiener-deconvolution. This interpretation of the $v$-update will guide our construction of a fully convolutional network later in Sec 3.3.

Algebraically, $(\mathbf{U}, \mathbf{U}^*)$ define an adjoint pair of warping operators, which we will refer to as "slicing" and "splatting" respectively [59]. In the case where there is injective motion relating a slice stack $\mathbf{f}^n$ to a 3D volume $\mathbf{v}$, we can obtain $\mathbf{f}^n$ by "slicing" $\mathbf{v}$ with $\mathbf{U}^n$ and obtain $\mathbf{v}$ by "splatting" $\mathbf{f}^n$ with $\mathbf{U}^{n*}$. Figure 3 illustrates the two adjoint warping operators in the discrete case when multi-linear interpolation weights are used for fractional slicing and splatting.

The $u$-update for the motion of the $k$th slice in a stack is

$$\mathbf{u}_k^n \leftarrow (\mathbf{S}^*\mathbf{V}_k^{n*}\mathbf{V}_k^n\mathbf{S})^{-1}\mathbf{S}^*\mathbf{V}_k^{n*}(\mathbf{H}_k\mathbf{v}^n - \mathbf{f}_k^n), \quad (8)$$

in which $\mathbf{V}_k^n = \mathbf{H}_k[\text{diag}(\mathbf{v}_x^n), \text{diag}(\mathbf{v}_y^n), \text{diag}(\mathbf{v}_z^n)]$, with the $\mathbf{v}_{\{x,y,z\}}^n$ denoting the respective spatial derivatives of the given 3D volume reconstructed in the $(n-1)$th iteration.

In practice, subject motion in an MR slice likely exceeds one voxel in magnitude but the linear model of warping (3) stays valid only if the slice is first blurred at the scale of the motion. Due to this, motion is often estimated on a pyramid in a coarse-to-fine manner, using the motion field estimated at coarser levels to initialize motion at finer ones [60]. MRI acquisitions also suffer from a bias field due to non-uniform main magnetic field and radiofrequency coils, introducing a global intensity shift between the unknown volume and the slices. A key challenge in optimization-based SVR methods is determining, at each scale, motion in the presence of bias and extreme slice artifacts. Mistaking either for motion can quickly lead to pitfalls of local minima in the optimization objective especially at coarse scales. However, it is difficult to hand-craft a reconstruction pipeline which can overcome image non-convexities due to sheer numbers of acquisition-specific design choices to be made (e.g., non-linear filters).

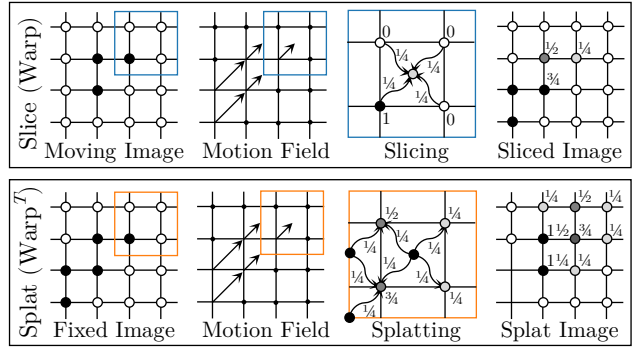

Figure 3: **Slicing and splatting.** In slicing (top row), the endpoints of motion vectors define coordinate locations in a moving image to pull data from and construct an image. In splatting (bottom), the same motion vector endpoints define coordinate locations in a fixed image to push data to. Multi-linear weights are used for fractional splatting and slicing, guaranteeing differentiability w.r.t grid data.

## 4. Fully Convolutional SVR

Using a FCN model for SVR enables us to bypass hand-crafted acquisition-specific reconstruction pipelines as well as numerical optimization. Rather than formulating SVR as the prediction of absolute slice coordinates in 3D space as is typically done in other learning-based methods [28–32], we cast SVR as the registration of some unobserved 3D volume to an observed stack of 2D slices. We train a FCN model to predict motion relating the slices to a 3D volume given only the slices as input, producing a 3D volume as a by-product of registration. Conceptually, this is similar to the problem of monocular depth estimation [40–43], where one's goal is to predict disparity (motion along the epipolar line) relating a single 2D image to the depth of the underlying 3D scene.

### 4.1. Neural Network Architecture

CNN architectures such as those used in semantic image segmentation can produce suboptimal outcomes for motion estimation tasks if the motion is large or images exhibit fine texture [33–39]. While fully convolutional networks such as the U-Net [54] are endowed with a multi-scale architecture and could theoretically estimate motion in a coarse-to-fine manner, using convolution kernels to parameterize warping (space-varying transforms) is inherently inefficient, both in terms of parametrization and computation; see, e.g., [33, 39].

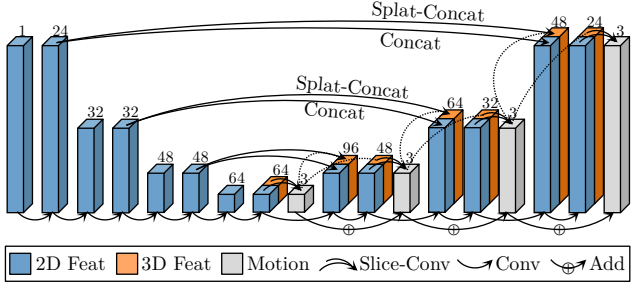

**Figure 4: Splat–Slice U-Net.** The stack of 2D slices goes through 2D feature extraction (downward path) and reconstruction (upward path) with 2D skip features. 3D volume features are reconstructed by splatting 2D skip features with previous motion to form 3D skip features. At each level, the sliced 3D feature volume and the slice feature stack are convolved jointly to extract residual slice motion.

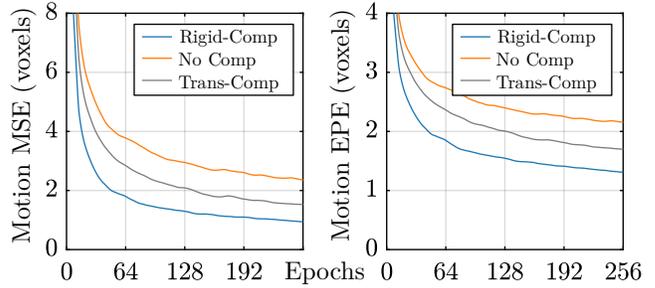

**Figure 5: Rigid motion-compensated loss.** Validation MSE and EPE (end-point error) [33] of the predicted motion shown. Our loss (blue), which compensates for offsets in global rigid motion, trains faster and attains higher final prediction accuracy than the MSE loss (orange) and the loss that compensates for translations only (gray).

Our network model (Figure 4) bears similarities to those used in pairwise image registration [28–32], which predict image motion by processing the image pair across multiple levels of convolutions. These network models are typically equipped with warping layers between adjacent levels such that only the residual motion needs to be estimated at every level. Notionally, the slice stack assumes the role of a fixed image while the 3D volume, which we seek to estimate, can be seen as a moving image. As discussed in Sec. 3.2, we can obtain the features of the 2D slices (fixed image) by slicing the features of the 3D volume with the estimated motion. To obtain the features of the 3D volume, we splat the features of the slice stack using the same motion estimate.

Note that in our case, only the fixed image (slice stack) is given, so the downward transform needs to process only the fixed image. The upward transform synthesizes fixed image features by unpooling lower-resolution ones, concatenating them with the skip features and processing them via a series of convolutions. The moving image (or 3D volume) features are synthesized in parallel by unpooling their low-resolution counterparts, concatenating them with the splattings of skip features and processing them via convolutions. The moving image features are finally sliced, concatenated with the fixed image features and processed using a series of convolutions to output residual slice motion, which is added to the lower-resolution motion that was used during splatting and slicing.

In terms of implementation, both the slice stack and the 3D volume features are seen as (B,C,D,H,W) tensors, where the D dimension acts as the slicing dimension in the case of slice stack features. 3D volume features are processed using convolution and pooling kernels of shape (3,3,3) and (2,2,2) respectively, while slice features are processed using kernels of shapes (1,3,3) and (1,2,2), respectively, so that each slice can be processed independently. Within the residual motion extraction stages, the first convolutions are performed with kernels of shape (4,3,3) and strides of (4,1,1) since slices are internally represented as 3D slabs with thickness of 4 voxels and the motion needs to be regularized across each slab. We handle different slice spacings in real data by scaling the 2D input slices to a 4:1 slab thickness to voxel spacing ratio.

### 4.2. Rigid Motion-Compensating Loss

Inexact positioning of subjects in MRI scanners typically introduces a global subject offset (rotations and translations) in the acquired slices. For reconstruction, however, we need only recover the motion of the slices relatively to each other ignoring global shifts. To ensure that we penalize prediction errors only in the relative slice motions, we compensate for any global rigid motion shift that may exist in our predicted motion stack before computing the training loss against the prediction target. Assuming an MSE training objective, we can express our motion-compensating training loss between the predicted $\mathbf{u}$ and ground truth $\mathbf{y} \in \mathbb{R}^{N \times 3}$ motion fields as

$$\mathcal{L}(\mathbf{u}, \mathbf{y}) = \min_{[\mathbf{R},\mathbf{t}]} \|\mathbf{u} + \mathbf{p} - (\mathbf{y}+\mathbf{p})\mathbf{R} - \mathbf{t}\|_F^2, \quad (9)$$

in which $\mathbf{R}$ represents a $3 \times 3$ rotation matrix, $\mathbf{t}$ represents a $1 \times 3$ translation vector, and $\mathbf{p} \in \mathbb{R}^{N \times 3}$ is a matrix of voxel coordinates. In words, this loss first performs a global rigid alignment between the point clouds $\mathbf{u} + \mathbf{p}$ and $\mathbf{y} + \mathbf{p}$, such that only the non-rigid motion component between the two point clouds is penalized. Loss $\mathcal{L}$ reduces to the usual MSE loss in the absence of global rotation $\mathbf{R}$ and translation $\mathbf{t}$.

To solve the optimization problem (9), we first minimize the unconstrained version of problem (9) over the space of all $[\mathbf{R}, \mathbf{t}] \in \mathbb{R}^{3 \times 4}$ to obtain a minimizer $[\mathbf{R}', \mathbf{t}^\star]$, then factor $\mathbf{R}'$ into its rotation and shear matrices as $\mathbf{R}' = \mathbf{R}^\star \mathbf{S}^\star$. From the singular value decomposition $\mathbf{R}' = \mathbf{U} \mathbf{\Sigma} \mathbf{V}^*$, we see that

$$\mathbf{R}^\star = \mathbf{U}\operatorname{sign}(\mathbf{\Sigma})\mathbf{V}^*, \qquad \mathbf{S}^\star = \mathbf{V}\operatorname{abs}(\mathbf{\Sigma})\mathbf{V}^*, \quad (10)$$

a factorization known as the polar decomposition [61]. The resulting rotation matrix $\mathbf{R}^\star$ is the minimum mean squared-error approximation of $\mathbf{R}'$, and consequently, the solution of the original rigid motion-constrained problem (9) is given by the pair $[\mathbf{R}^\star, \mathbf{t}^\star]$. The loss $\mathcal{L}(\mathbf{u}, \mathbf{y})$ is then backpropagated and the weights of our SVR network model updated. Figure

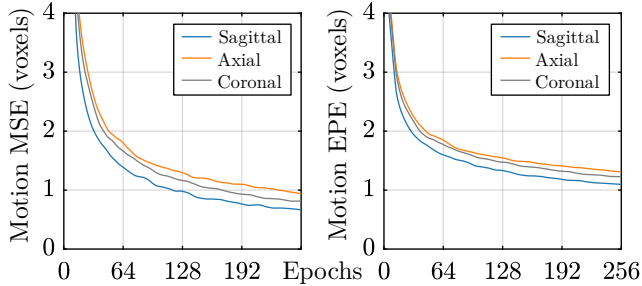

**Figure 6: Accuracy of predicted motion.** Slicing direction has an impact on the accuracy of predicted motion. We plot the validation MSE (left), and EPE (end-point error, right) of the predicted motion on adult brain slices for sagittal, axial, and coronal acquisitions.

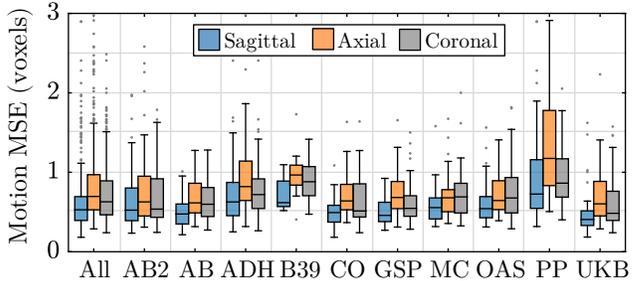

**Figure 7: Motion accuracy across datasets.** We plot the MSE of the motion stacks predicted on 500 held-out slice stacks (1mm$^3$, see text). Motion prediction is accurate to < 1mm on average. Sagittal and coronal motion predictions tend to be more accurate than axial.

5 plots the validation curves for our network model trained using motion-compensated and regular losses, showing that our rigid compensating loss is key to improving prediction.

### 4.3. Interpolating Reconstructions

In general, a 3D volume reconstructed from a single slice stack can contain holes if regions of the underlying volume are missed in all slice acquisitions due to subject motion. In comparison with natural images, MR scans typically exhibit a limited diversity in imagery and have a strong prior on the distribution of underlying 3D volumes. This justifies the use of interpolation techniques to fill holes in reconstructions to satisfy regularity constraints (e.g., on pathology). We turn to a fully convolutional interpolation model trained in advance on (true reconstruction, volume) pairs. True reconstructions (with holes) can be obtained readily at train time by warping slice stacks with ground truth slice motion. While the search for the best interpolation network model is not the focus of this work, we find that a standard U-Net model trained with true reconstructions as the target and using the mean squared error loss produces good results. Figure 2 (right) shows the training procedure for our fully convolutional interpolator.

## 5. Experimental Results

We run extensive experiments on both adult MRI stacks (for which ground truth MR volumes are available) and fetal ones, for which only reference volumes are given. For each experiment, we train a SVR network in a supervised fashion on paired (slice stack, motion stack) data and an interpolator network on paired (slice–splat volume, underlying volume) data (Sec. 4.3). The data augmentation hyperparameters and training details are given in Appendix A. We use the mean squared error (MSE) and the average end-point error (EPE) [33] to measure prediction error; see Appendix A.

### 5.1. Single-Stack SVR of Adult Brains

SVR experiments on synthetic sliced MRI volumes can reveal qualitative and quantitative characteristics of a SVR method by facilitating comparison across the reconstructed and the ground truth 3D MR volumes. We curate 1100 adult brain MR scans from ABIDE, -2 [62], ADHD [63], COBRE [64], GSP [65], MCIC [66], OASIS [67], PPMI [68], UKB [69] and B39 [70], and split the scans into 1000 training and 100 validation examples. All scans have a $256^3$ volume with 1mm$^3$ resolution. The scans are aligned to the Talairach atlas [71] to facilitate slicing along the sagittal, axial and coronal directions; see Appendix A for details on training and sample generation.

**Slicing Directions and Reconstruction.** Single-stack SVR can achieve different reconstruction accuracy depending on the direction of slicing (sagittal, axial and coronal). To study the effect of slice direction on the accuracy of the predicted motion, we train a separate network for each slice direction and plot in Figure 6, the validation MSE and Endpoint Error of the motion predicted by each network. The final validation error attained is the lowest for sagittal stacks at 0.89 (MSE) and the highest for axial stacks at 1.33 (MSE). Such a slight increase in the prediction error for axial stacks most likely stems from fewer "corner" features to track and align in axial slices, leading to a less precise alignment.

**Motion Prediction Error.** We plot distributions of the per-subject errors of the predicted motion stacks across 500 held-out scans in Figure 7. As expected from the validation error curves seen in Figure 6, sagittal predictions have the lowest median motion MSE and the tightest interquartile range. On average, our predicted motion is accurate to less than a voxel although the PPMI dataset exhibits slightly higher errors.

**Reconstruction and Interpolation.** Figure 8 visualizes the 3D reconstructions obtained using the predicted motion. We align all reconstructions back to the Talairach atlas [71] for comparison across different reconstructions. We also include reconstructions with the missing intensity data interpolated using our interpolator described in Sec. 4.3. The splat results show that our network predicts linear motion fields without the need for additional linear projections. The interpolation removes holes as well as slicing artifacts that interfere with downstream tasks such as brain morphometry. Larger holes are seen when splattings are viewed along the slicing axis.

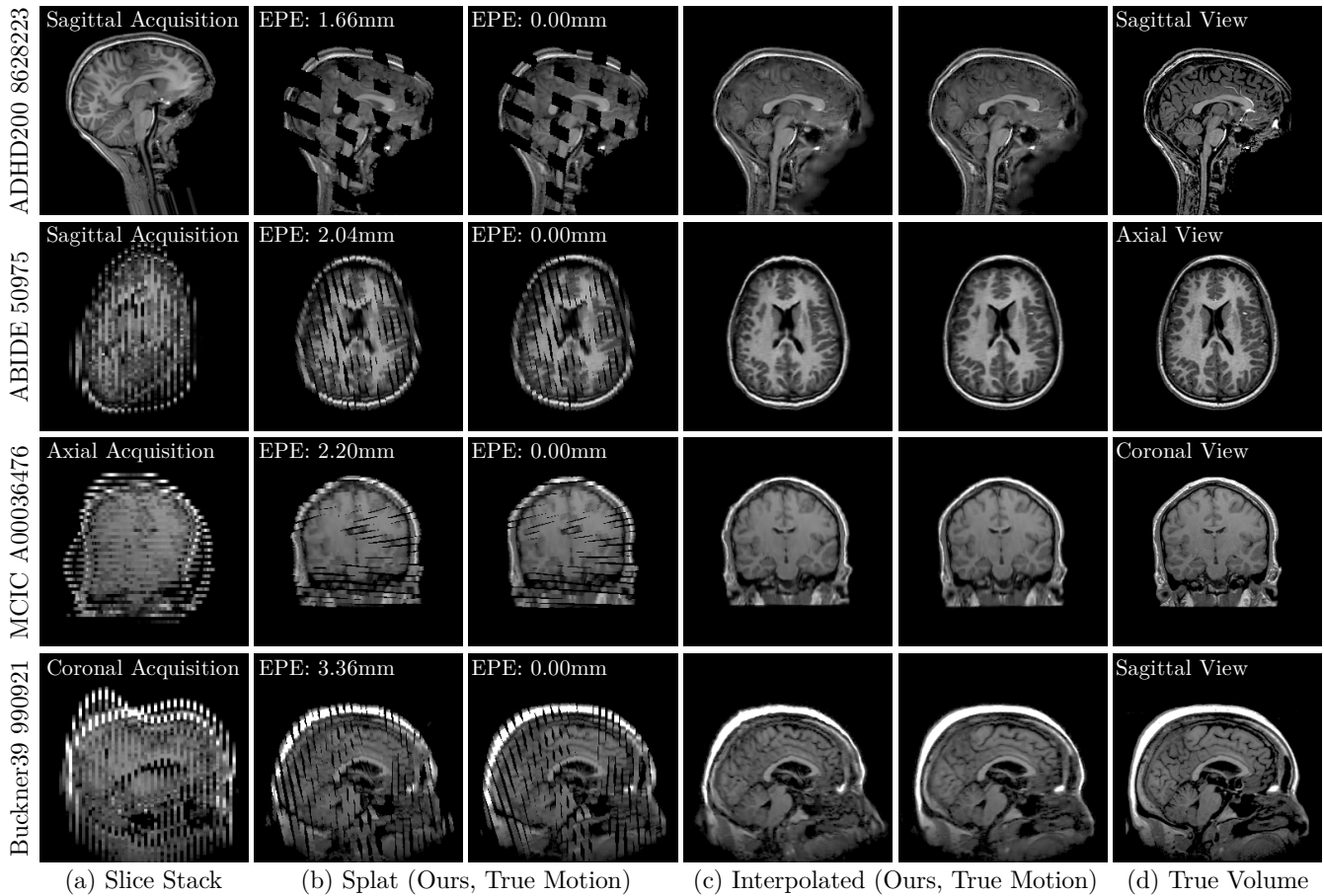

(a) Slice Stack    (b) Splat (Ours, True Motion)    (c) Interpolated (Ours, True Motion)    (d) True Volume

**Figure 8: SVR of adult brain scans.** We visualize our SVR results on sagittal (rows 1–2), axial (row 3), and coronal (last row) slice stacks synthesized using random slice motion (a). Using the motion stack predicted by our network, we splat slice data to reconstruct the underlying 3D volume (b). Using our pre-trained interpolator, we then interpolate the missing intensities (holes) in our reconstruction (c). The result is similar to the true 3D volume (d). We additionally visualize in (b) and (c) the splat and interpolated results when the true motion is used.

### 5.2. Single-Stack SVR of Fetal Brains

Here, we curate 98 T2w fetal brain atlases and reference reconstructions across the CRL Fetal Atlases [72] and FeTA [73] (0.8mm³, aligned to age-matched CRL atlases) and split them into 86 training and 12 validation FeTA examples. We use two real T2w slice stacks from MIAL [16] for test. In the case of fetal SVR, we train a single splat–slice network on all possible slicing directions by applying a random rotation (Euler angles between ±180º) to our training examples. We compare our method with state-of-the-art SVRnet [20] and SVoRT [25]; see Appendix A. Optimization-based methods do not work on single stacks and are not compared against.

**3D Reconstruction and Interpolation.** Figure 9 visualizes the reconstructions obtained using our predicted motion. We align FeTA reconstructions to their gestational age-matched atlas [72] for comparison across different methods. SVRnet [20] completely fails to align slices in most of the cases and SVoRT (v2) introduces large slice misalignments, leading to suboptimal 3D reconstruction with apparent distortion. (The SVoRT and SVRnet reconstructions are processed using our interpolation.) Our reconstructions are more faithful to the underlying 3D volumes up to ambiguities from slicing. The last two rows visualize reconstruction of the two real MIAL acquisitions for which no ground truth exists and motion is smaller. Unlike SVRnet, the proposed method generalizes to real acquisitions even when the network model is trained on synthetic slice stacks with different motion and acquisition parameters, e.g., PSF. See Appendix B for more examples.

**Quantitative Results.** Table 1 lists the prediction accuracy of different methods on validation subjects with 4 folds. In addition to motion EPE, we report the slice PSNR, obtained by slicing the true volume with the predicted motion, and the PSNR of the final reconstruction. The three-stack results are from [25], which lists the anchor point errors (APE). In the single-stack case, our method reduces error in the predicted motion by 77.6% and 44.6% relative to the SVRnet [20] and SVoRT (v2) predictions, respectively. The slice and volume PSNR have also improved but unlike the EPE, these metrics depend on the PSF of acquisition and reconstruction and do not definitively measure alignment accuracy.

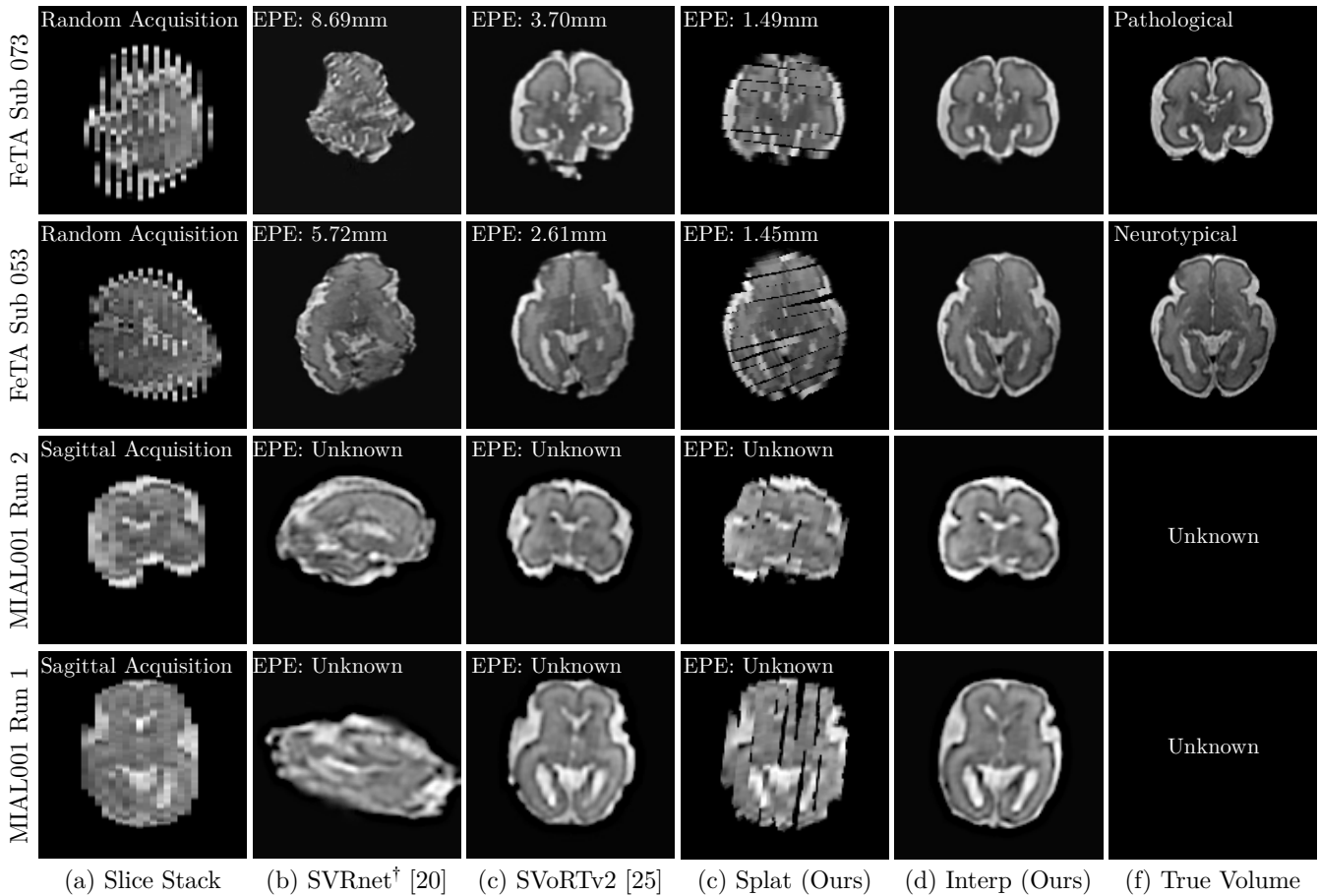

(a) Slice Stack    (b) SVRnet[†] [20]    (c) SVoRTv2 [25]    (c) Splat (Ours)    (d) Interp (Ours)    (f) True Volume

**Figure 9: Single-stack fetal SVR.** We visualize the SVR results on validation subjects from the FeTA dataset [73] and two real acquisitions from MIAL [16]. Zoomed in 4x for better visibility. Our results closely resemble the ground truth volumes while SVoRTv2 and SVRnet reconstructions (with our interpolation) exhibit spatial distortion from inaccurate slice alignment. [†]Our implementation; see Appendix A.

| | Method | APE/Motion EPE (mm) | Slice PSNR (dB) | Volume PSNR (dB) | Time (sec) |
|---|---|---|---|---|---|
| 3 Stacks | SVRnet* [20] | 12.82±5.69 | 20.53±1.62 | 19.54±1.52 | – |
| | PlaneInVol* [24] | 12.49±6.73 | 19.96±1.73 | 18.98±1.62 | – |
| | SVoRT* [25] | 4.35±0.90 | 25.26±1.86 | 23.32±1.42 | – |
| 1 Stack | SVRnet[†] [20] | 8.08±2.35 | 13.46±1.56 | 16.03±1.28 | **0.011s** |
| | SVoRT (v2) [25] | 3.27±0.71 | 20.77±1.28 | 18.49±1.63 | 0.142s |
| | Proposed (Ours) | **1.81±0.40** | **23.69±1.39** | **23.43±1.40** | 0.224s |

**Table 1: Validation accuracy.** We list the motion and 3D volume reconstruction accuracy of different methods. *Results taken from [25]. Timed on RTX8000. [†]Our implementation; see Appendix A.

## 6. Discussion

**Limitations.** Our fetal SVR network is currently trained on automatically segmented brain slices. Training with original unsegmented slice stacks can improve the robustness of our approach against imperfectly generated segmentations.

**Future Work.** Given that our approach predicts dense slice motion for reconstruction, we plan to extend our approach to problems where the acquired slices can undergo deformable motion—e.g. in fetal torso or placental reconstruction [18]—and cannot be tackled using rigid motion SVR. Also, better interpolation networks, such as those that are equivariant to rigid motion, can further improve reconstruction; see [74].

## 7. Conclusion

In brain imaging, slice-to-volume reconstruction (SVR) is an important computational technique for the imaging of subjects whose motion cannot easily be controlled. Yet, the majority of SVR techniques previously proposed do not reap the benefits of FCN models that have gradually become the mainstay in closely related tasks, such as image registration and segmentation. The few learning-based SVR techniques resort to pretrained ImageNet classification and even vision transformer models. Not only does our work fill the gap in the SVR method development timeline but it also shows that a FCN model can produce state-of-the-art SVR outcomes.

**Acknowledgments.** SIY thanks Margherita Firenze for her thought-provoking questions. Work primarily supported by the NIH grant K99AG081493 with additional support from grants RF1MH123195, R01AG064027 and R21AG082082.

# References

[1] Mitsuhiro Nishida et al. Detailed semiautomated MRI based morphometry of the neonatal brain: Preliminary results. *NeuroImage*, 32(3):1041–1049, 2006.

[2] Chao J. Liu et al. Quantification of volumetric morphometry and optical property in the cortex of human cerebellum at micrometer resolution. *NeuroImage*, 244:118627, 2021.

[3] Maxwell L. Elliott et al. Brain morphometry in older adults with and without dementia using extremely rapid structural scans. *NeuroImage*, 276:120173, 2023.

[4] Bruce Fischl. FreeSurfer. *NeuroImage*, 62(2):774–781, 2012.

[5] Amy Zhao, Guha Balakrishnan, Fredo Durand, John V. Guttag, and Adrian V. Dalca. Data augmentation using learned transformations for one-shot medical image segmentation. In *Proc. CVPR*, 2019.

[6] Lilla Zöllei, Juan Eugenio Iglesias, Yangming Ou, P. Ellen Grant, and Bruce Fischl. Infant FreeSurfer: An automated segmentation and surface extraction pipeline for T1-weighted neuroimaging data of infants 0–2 years. *NeuroImage*, 218:116946, 2020.

[7] Alan C. Evans, Andrew L. Janke, D. Louis Collins, and Sylvain Baillet. Brain templates and atlases. *NeuroImage*, 62(2):911–922, 2012.

[8] Shadab Khan et al. Fetal brain growth portrayed by a spatiotemporal diffusion tensor MRI atlas computed from in utero images. *NeuroImage*, 185:593–608, 2019.

[9] Adrià Casamitjana, and Juan Eugenio Iglesias. High-resolution atlasing and segmentation of the subcortex: Review and perspective on challenges and opportunities created by machine learning. *NeuroImage*, 263:119616, 2022.

[10] Shuzhou Jiang, Hui Xue, Alan Glover, Mary Rutherford, Daniel Rueckert, and Joseph V. Hajnal. MRI of moving subjects using multislice snapshot images with volume reconstruction (SVR): Application to fetal, neonatal, and adult brain studies. *IEEE Trans. Med. Imaging*, 26(7):967–980, 2007.

[11] Ali Gholipour, Judy A. Estroff, and Simon K. Warfield. Robust super-resolution volume reconstruction from slice acquisitions: Application to fetal brain MRI. *IEEE Trans. Med. Imaging*, 29(10):1739–1758, 2010.

[12] Kio Kim, Piotr A. Habas, Francois Rousseau, Orit A. Glenn, Anthony J. Barkovich, and Colin Studholme. Intersection based motion correction of multislice MRI for 3-D in utero fetal brain image formation. *IEEE Trans. Med. Imaging*, 29(1):146–158, 2010.

[13] Maria Kuklisova-Murgasova, Gerardine Quaghebeur, Mary A. Rutherford, Joseph V. Hajnal, and Julia A. Schnabel. Reconstruction of fetal brain MRI with intensity matching and complete outlier removal. *Med. Image Anal.*, 16(8):1550–1564, 2012.

[14] Bernhard Kainz et al. Fast volume reconstruction from motion corrupted stacks of 2D slices. *IEEE Trans. Med. Imaging*, 34(9):1901–1913, 2015.

[15] Amir Alansary et al. PVR: Patch-to-Volume Reconstruction for large area motion correction of fetal MRI. *IEEE Trans. Med. Imaging*, 36(10):2031–2044, 2017.

[16] Sébastien Tourbier et al. Automated template-based brain localization and extraction for fetal brain MRI reconstruction. *NeuroImage*, 155:460–472, 2017.

[17] Michael Ebner et al. An automated framework for localization, segmentation and super-resolution reconstruction of fetal brain MRI. *NeuroImage*, 206:116324, 2020.

[18] Alena Uus, Tong Zhang, Laurence H. Jackson, Thomas A. Roberts, Mary A. Rutherford, Joseph V. Hajnal, and Maria Deprez. Deformable slice-to-volume registration for motion correction of fetal body and placenta MRI. *IEEE Trans. Med. Imaging*, 39(9):2750–2759, 2020.

[19] Francois Rousseau, Orit A. Glenn, Bistra Iordanova, Claudia Rodriguez-Carranza, Daniel B. Vigneron, James A. Barkovich, and Colin Studholme. Registration-based approach for reconstruction of high-resolution in utero fetal MR brain images. *Acad. Radiol.*, 13(9):1072–1081, 2006.

[20] Benjamin Hou et al. 3-D Reconstruction in canonical co-ordinate space from arbitrarily oriented 2-D Images. *IEEE Trans. Med. Imaging*, 37(8):1737–1750, 2018.

[21] Benjamin Hou et al. Computing CNN loss and gradients for pose estimation with Riemannian geometry. In *Proc. MICCAI*, 2018.

[22] Seyed Sadegh Mohseni Salehi, Shadab Khan, Deniz Erdogmus, and Ali Gholipour. Real-time deep pose estimation with geodesic loss for image-to-template rigid registration. *IEEE Trans. Med. Imaging*, 38(2):470–481, 2019.

[23] Yuchen Pei, Lisheng Wang, Fenqiang Zhao, Tao Zhong, Lufan Liao, Dinggang Shen, and Gang Li. Anatomy-guided convolutional neural network for motion correction in fetal brain MRI. In *Proc. MLMI*, 2020.

[24] Pak-Hei Yeung, Moska Aliasi, Aris T. Papageorghiou, Monique Haak, Weidi Xie, and Ana I. L. Namburete. Learning to map 2D ultrasound images into 3D space with minimal human annotation. *Med. Image Anal.*, 70:101998, 2021.

[25] Junshen Xu, Daniel Moyer, P. Ellen Grant, Polina Golland, Juan Eugenio Iglesias, and Elfar Adalsteinsson. SVoRT: Iterative transformer for slice-to-volume registration in fetal brain MRI. In *Proc. MICCAI*, 2022.

[26] Elise Turk et al. Functional connectome of the fetal brain. *J. Neurosci.*, 39(49):9716–9724, 2019.

[27] Daniel Sobotka et al. Motion correction and volumetric reconstruction for fetal functional magnetic resonance imaging data. *NeuroImage*, 255:119213, 2022.

[28] Xiaohuan Cao, Jianhua Yang, Jun Zhang, Dong Nie, Minjeong Kim, Qian Wang, and Dinggang Shen. Deformable image registration based on similarity-steered CNN regression. In *Proc. MICCAI*, 2017.

[29] Guha Balakrishnan, Amy Zhao, Mert R. Sabuncu, John Guttag, and Adrian V. Dalca. An unsupervised learning model for deformable medical image registration. In *Proc. CVPR*, 2018.

[30] Tony C. W. Mok, and Albert C. S. Chung. Fast symmetric diffeomorphic image registration with convolutional neural networks. In *Proc. CVPR*, 2020.

[31] Sean I. Young, Yaël Balbastre, Adrian V. Dalca, William M.

Wells, Juan Eugenio Iglesias, and Bruce Fischl. SuperWarp: Supervised learning and warping on U-Net for invariant subvoxel-precise registration. In *Proc. WBIR*, 2022.

[32] Miao Kang, Xiaojun Hu, Weilin Huang, Matthew R. Scott, and Mauricio Reyes. Dual-stream pyramid registration network. *Med. Image Anal.*, 78:102379, 2022.

[33] Alexey Dosovitskiy et al. FlowNet: Learning Optical Flow With Convolutional Networks. In *Proc. ICCV*, 2015.

[34] Jason J. Yu, Adam W. Harley, and Konstantinos G. Derpanis. Back to basics: Unsupervised learning of optical flow via brightness constancy and motion smoothness. In *Proc. ECCV*, 2016.

[35] Eddy Ilg, Nikolaus Mayer, Tonmoy Saikia, Margret Keuper, Alexey Dosovitskiy, and Thomas Brox. FlowNet 2.0: Evolution of optical flow estimation with deep networks. In *Proc. CVPR*, 2017.

[36] Deqing Sun, Xiaodong Yang, Ming-Yu Liu, and Jan Kautz. PWC-Net: CNNs for optical flow using pyramid, warping, and cost volume. In *Proc. CVPR*, 2018.

[37] Anurag Ranjan, and Michael J. Black. Optical flow estimation using a spatial pyramid network. In *Proc. CVPR*, 2017.

[38] Pengpeng Liu, Michael Lyu, Irwin King, and Jia Xu. SelFlow: Self-supervised learning of optical flow. In *Proc. CVPR*, 2019.

[39] Junhwa Hur, and Stefan Roth. Iterative residual refinement for joint optical flow and occlusion estimation. In *Proc. CVPR*, 2019.

[40] Fayao Liu, Chunhua Shen, and Guosheng Lin. Deep convolutional neural fields for depth estimation from a single image. In *Proc. CVPR*, 2015.

[41] Jae-Han Lee, and Chang-Su Kim. Monocular depth estimation using relative depth maps. In *Proc. CVPR*, 2019.

[42] Clement Godard, Oisin Mac Aodha, Michael Firman, and Gabriel J. Brostow. Digging into self-supervised monocular depth estimation. In *Proc. ICCV*, 2019.

[43] Shariq Farooq Bhat, Ibraheem Alhashim, and Peter Wonka. AdaBins: Depth estimation using adaptive bins. In *Proc. CVPR*, 2021.

[44] Jia Deng, Wei Dong, Richard Socher, Li-Jia Li, Kai Li, and Li Fei-Fei. ImageNet: a large-scale hierarchical image database. In *Proc. CVPR*, 2009.

[45] Alexey Dosovitskiy et al. An image is worth 16x16 words: Transformers for image recognition at scale. In *Proc. ICLR*, 2020.

[46] Anurag Arnab, Mostafa Dehghani, Georg Heigold, Chen Sun, Mario Lučić, and Cordelia Schmid. ViViT: A video vision transformer. In *Proc. ICCV*, 2021.

[47] Ze Liu et al. Swin Transformer: Hierarchical vision transformer using shifted windows. In *Proc. ICCV*, 2021.

[48] Jeremie Anquez, Elsa D. Angelini, and Isabelle Bloch. Automatic segmentation of head structures on fetal MRI. In *Proc. ISBI*, 2009.

[49] Bernhard Kainz, Kevin Keraudren, Vanessa Kyriakopoulou, Mary Rutherford, Joseph V. Hajnal, and Daniel Rueckert. Fast fully automatic brain detection in fetal MRI using dense rotation invariant image descriptors. In *Proc. ISBI*, 2014.

[50] Kevin Keraudren, Vanessa Kyriakopoulou, Mary Rutherford, Joseph V. Hajnal, and Daniel Rueckert. Localisation of the brain in fetal MRI using bundled SIFT features. In *Proc. MICCAI*, 2013.

[51] Kevin Keraudren et al. Automated fetal brain segmentation from 2D MRI slices for motion correction. *NeuroImage*, 101:633–643, 2014.

[52] Martin Rajchl et al. Learning under distributed weak supervision, http://arxiv.org/abs/1606.01100, 2016.

[53] Seyed Sadegh Mohseni Salehi et al. Real-time automatic fetal brain extraction in fetal MRI by deep learning. In *Proc. ISBI*, 2018.

[54] Olaf Ronneberger, Philipp Fischer, and Thomas Brox. U-Net: convolutional networks for biomedical image segmentation. In *Proc. MICCAI*, 2015.

[55] Junshen Xu, Daniel Moyer, Borjan Gagoski, Juan Eugenio Iglesias, P. Ellen Grant, Polina Golland, and Elfar Adalsteinsson. NeSVoR: Implicit neural representation for slice-to-volume reconstruction in MRI. *IEEE Trans. Med. Imaging*, 42(6):1707–1719, 2023.

[56] Ben Mildenhall, Pratul P. Srinivasan, Matthew Tancik, Jonathan T. Barron, Ravi Ramamoorthi, and Ren Ng. NeRF: Representing scenes as neural radiance fields for view synthesis, In *Proc. ECCV*, 2020.

[57] Sahar N. Saleem. Fetal MRI: An approach to practice: A review. *J. Adv. Res.*, 5(5):507–523, 2014.

[58] Hua-mei Chen, and P.K. Varshney. Mutual information-based CT-MR brain image registration using generalized partial volume joint histogram estimation. *IEEE Trans. Med. Imaging*, 22(9):1111–1119, 2003.

[59] Andrew Adams, Jongmin Baek, and Myers Abraham Davis. Fast high-dimensional filtering using the permutohedral lattice. *Comput. Graph. Forum*, 29(2):753–762, 2010.

[60] Nils Papenberg, Andrés Bruhn, Thomas Brox, Stephan Didas, and Joachim Weickert. Highly accurate optic flow computation with theoretically justified warping. *Int. J. Comput. Vis.*, 67(2):141–158, 2006.

[61] Nicholas J. Higham. Computing the polar decomposition—with applications. *SIAM J. Sci. Stat. Comput.*, 7(4):1160–1174, 1986.

[62] A. Di Martino et al. The autism brain imaging data exchange: towards a large-scale evaluation of the intrinsic brain architecture in autism. *Mol. Psychiatry*, 19(6):659–667, 2014.

[63] Michael Milham, Damien Fair, Maarten Mennes, and Stewart Mostofsky. The ADHD-200 consortium: a model to advance the translational potential of neuroimaging in clinical neuroscience. *Front. Syst. Neurosci.*, 6, 2012.

[64] Vince Calhoun, Jing Sui, Kent Kiehl, Jessica Turner, Elena Allen, and Godfrey Pearlson. Exploring the psychosis functional connectome: aberrant intrinsic networks in schizophrenia and bipolar disorder. *Front. Psychiatry*, 2, 2012.

[65] Avram J. Holmes et al. Brain Genomics Superstruct Project initial data release with structural, functional, and behavioral measures. *Sci. Data*, 2(1):150031, 2015.

[66] Randy L. Gollub et al. The MCIC Collection: a shared repository of multi-modal, multi-site brain image data from a clinical investigation of schizophrenia. *Neuroinformatics*,

11(3):367–388, 2013.

[67] Daniel S. Marcus, Tracy H. Wang, Jamie Parker, John G. Csernansky, John C. Morris, and Randy L. Buckner. Open Access Series of Imaging Studies (OASIS): Cross-sectional MRI data in young, middle aged, nondemented, and demented older adults. *J. Cogn. Neurosci.*, 19(9):1498–1507, 2007.

[68] Kenneth Marek et al. The Parkinson Progression Marker Initiative (PPMI). *Prog. Neurobiol.*, 95(4):629–635, 2011.

[69] Cathie Sudlow et al. UK Biobank: an open access resource for identifying the causes of a wide range of complex diseases of middle and old age. *PLOS Med.*, 12(3):e1001779, 2015.

[70] Bruce Fischl et al. Whole brain segmentation: Automated labeling of neuroanatomical structures in the human brain. *Neuron*, 33(3):341–355, 2002.

[71] J. Talairach, and Pierre Tournoux. Co-planar stereotaxic atlas of the human brain: 3-dimensional proportional system : an approach to cerebral imaging. G. Thieme ; Thieme Medical Publishers, Stuttgart, New York 1988.

[72] Ali Gholipour et al. A normative spatiotemporal MRI atlas of the fetal brain for automatic segmentation and analysis of early brain growth. *Sci. Rep.*, 7(1):476, 2017.

[73] Kelly Payette et al. An automatic multi-tissue human fetal brain segmentation benchmark using the Fetal Tissue Annotation Dataset. *Sci. Data*, 8(1):167, 2021.

[74] Dongdong Chen, Mike Davies, Matthias J. Ehrhardt, Carola-Bibiane Schönlieb, Ferdia Sherry, and Julián Tachella. Imaging with equivarient deep learning: from unrolled network design to fully unsupervised learning. *IEEE Signal Process. Mag.*, 40(1):134–147, 2023.

## A. Training and Validation

### A.1 Training Details

For training, we prepare paired (slice stack, motion stack) and (splatted volume, ground truth volume) training data by applying a random isotropic zoom (±26 voxels), horizontal flip, rotation (Euler angles ±20º for adult brains and ±180º for fetal) and translation (±13 voxels) to the registered $256^3$ reference volume (1mm³ acquisition, Talairach for adult and 0.8mm³ FeTA reconstruction, CRL for fetal) to first simulate the diversity in scanned subjects and various poses that they may be scanned under. To simulate slicing, we apply random rotations (Euler angles ±20º) and translations (±26 voxels) that vary smoothly across slices by randomly sampling 32 to 64 rotations and translations from the ranges shown above and smoothly interpolating them using cubic B splines. We interleave the two halves of motion trajectories to simulate two-shot sequences typically used in 2D MRI acquisitions.

Slice acquisition is simulated by blurring the slices using a boxcar PSF (four voxels wide) along the slicing direction (sagittal, axial or coronal for adult brains, and sagittal in the case of fetal, where the ±180º Euler angles mean that slices are acquired along random directions anyway) and sampling every fourth slice along the same axis. Slice intensities are manipulated by applying Gaussian noise with noise standard deviation $\sigma = 0.01$ and gamma augmentation with exponent $\gamma \in [0.9, 1]$. Finally, the acquired slices are replicated along the slicing direction by a factor of four again so that the slice stacks have isotropic in- and through-plane resolutions. We subsample the slice stacks by a factor of two (i.e., to 2mm³ or 1.6mm³) to train the slice motion prediction network. No subsampling is performed when training the interpolator.

We train our slice motion and interpolation networks for 256,000 steps using ADAM, with an initial learning rate of $10^{-4}$, which is reduced to 0 with poly scheduling (exponent of 0.9), weight decay of 0 and momentum of 0.90. We mask out background voxels when computing the training loss. In the case of fetal SVR, we over-sample the FeTA portion of the training data fivefold, and the CRL portion tenfold. We simply pick the model of the last epoch for model selection but still monitor validation metrics for potential overfitting.

### A.2 Validation Metrics

Similar to our training loss (9), all our validation metrics compensate for any global rigid motion offset may that exist between predicted and true slice motion; see (9). In addition to the MSE $\mathcal{L}_{\text{MSE}}(\mathbf{u}, \mathbf{y}) = (1/N)\|\mathbf{u} - \mathbf{y}\|_F^2$ of the predicted slice motion $\mathbf{u} \in \mathbb{R}^{N \times 3}$ w.r.t. true motion $\mathbf{y} \in \mathbb{R}^{N \times 3}$, we use the average end-point error (EPE) metric [33], defined as

$$\mathcal{L}_{\text{EPE}}(\mathbf{u}, \mathbf{y}) = (1/N)\|\mathbf{u} - \mathbf{y}\|_{2,1}, \quad (A1)$$

that is, the mean Euclidean distance between the end-points of two slice motion fields (both metrics shown here without rigid compensation for clarity). In previous work [20, 25], a similar metric is proposed to measure the average Euclidean distance between the predicted and true slice positions at the anchor points of the slices (anchor point error, APE):

$$\mathcal{L}_{\text{APE}}(\mathbf{u}, \mathbf{y}) = (1/3)\|\mathbf{u}_{\{0,1,2\}} - \mathbf{y}_{\{0,1,2\}}\|_{2,1}, \quad (A2)$$

in which $\mathbf{u}_{\{0,1,2\}}$ and $\mathbf{y}_{\{0,1,2\}}$ denote the position vectors at the anchor points of the grids that define the voxel locations of the respective slices in 3D space. Typically, anchor points are assumed to be at the center, bottom left and bottom right corners of a given slice. Assuming that slices are undergoing rigid motion, the APE is equivalent to the EPE averaged on the right-triangular region formed by the three anchor points on the reference slice.

### A.3 SVRnet Model Implementation

For reproducibility, we port the original TensorFlow 1.13 implementation of SVRnet [20] with an Inception backbone to PyTorch 1.13, where we use a ResNet-34 backbone and a prediction head consisting of a $512 \times 9$ dense layer, which predicts the slice position vectors at three anchor points. We find that 2D batch normalization based on collected statistics does not perform well at test time and opt to normalize each example based on the statistics of each slice stack. We use subsampled slices of size $128 \times 128$ pixels and interpolate the predicted anchor point position vectors to a linear motion field and subtract the slice voxel coordinates to output slice motion. We train this implementation of SVRnet using the regular MSE loss on output slice motion field. We initialize the model with the torchvision ImageNet1K_v1 weights.

### A.4 SVoRT Model Configuration

For comparison with SVoRT (v2), we use model weights provided by Xu et al. [25] on their repository, configure the model to use one slice stack with a slice gap of 3.2mm, and optimize the reconstruction PSF (slice thickness of 1.6mm) for validation accuracy (i.e., average motion end-point error) on the 12 FeTA validation subjects using an exponential grid search. We fixed the stack positional encoding of SVoRT to 0 (rather than a random integer) for reproducible results. We convert SVoRT's transform output to dense motion fields to compute the motion MSE and EPE for validation accuracy.

## B. Additional SVR Results

Here, we provide additional SVR results on our adult and fetal datasets. Figure B1 visualizes reconstructions of adult brain MR volumes for three of our validation subjects, with all three orthogonal views shown for completeness. Figure B2 similarly visualizes fetal reconstructions. We include the corresponding SVoRTv2 results for comparison, noting that SVRnet reconstructions are garbled in many cases (see first row of Figure 9) and are less meaningful to compare against.

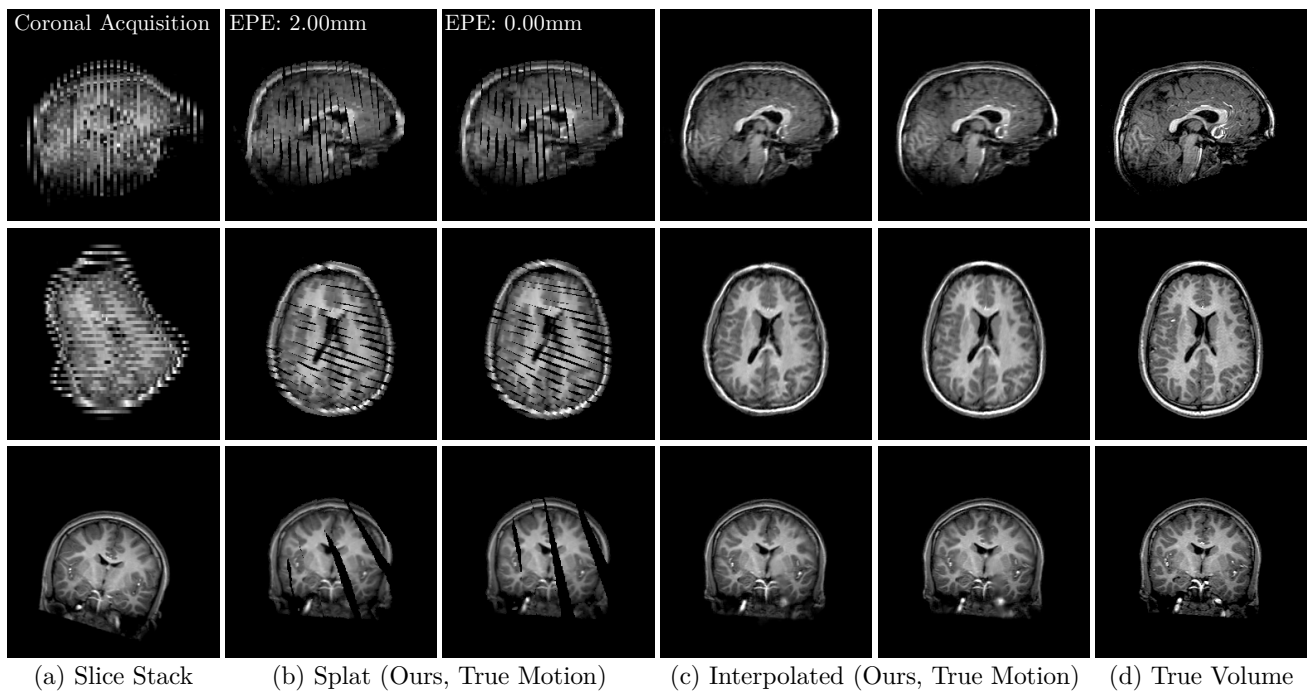

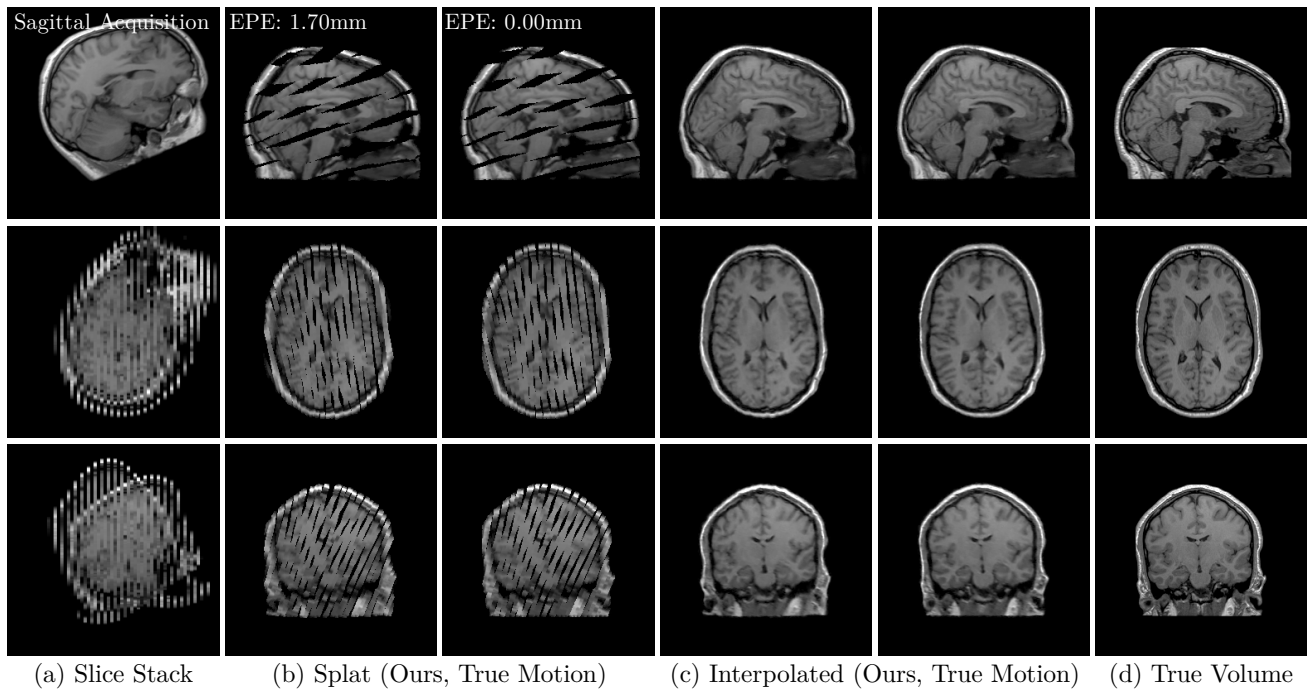

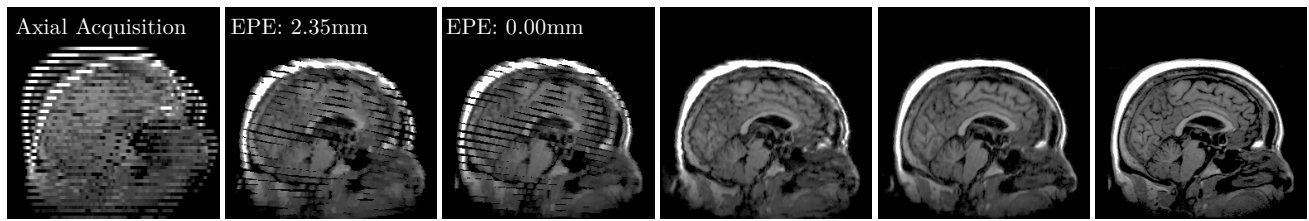

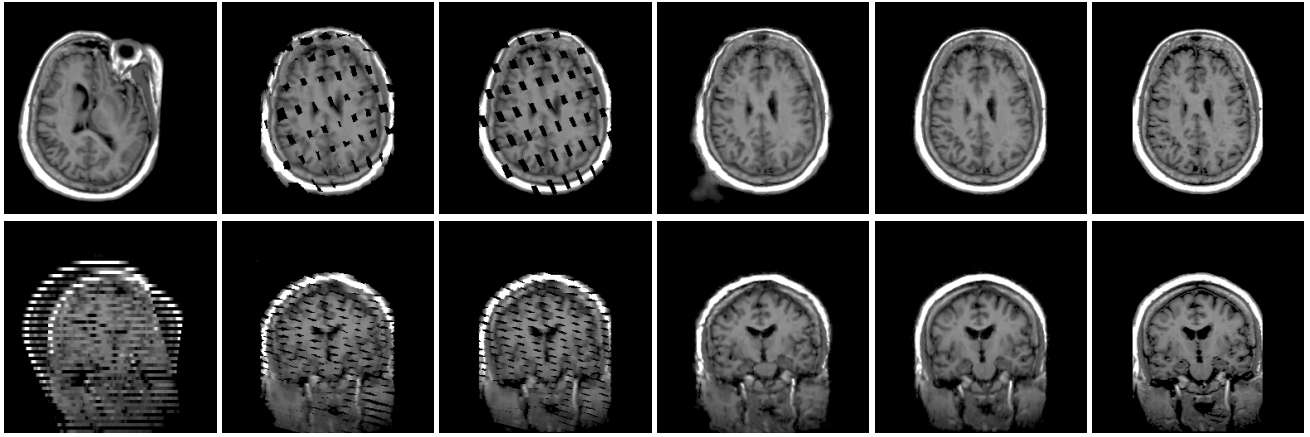

(a) Slice Stack    (b) Splat (Ours, True Motion)    (c) Interpolated (Ours, True Motion)    (d) True Volume

**Figure B1: SVR of adult brain scans.** We visualize our SVR results on slice stacks synthesized using random slice motion (a). Using the predicted motion stack, we splat slice data to reconstruct the underlying 3D volume (b). We interpolate the missing intensities (holes) in our reconstruction (c). We additionally visualize in (b) and (c) splat and interpolated results obtained when the true motion stack is used.

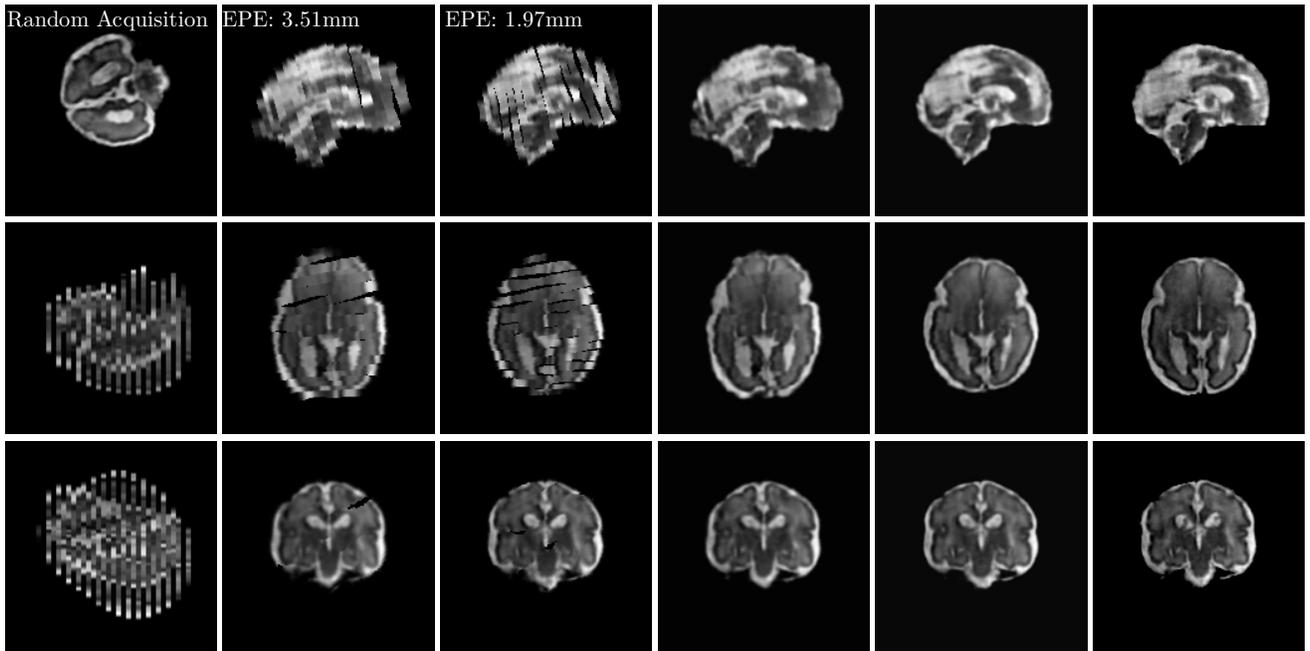

(a) Slice Stack    (b) Splat (SVoRTv2, Ours)    (c) Interp (SVoRTv2, Ours)    (d) True Volume

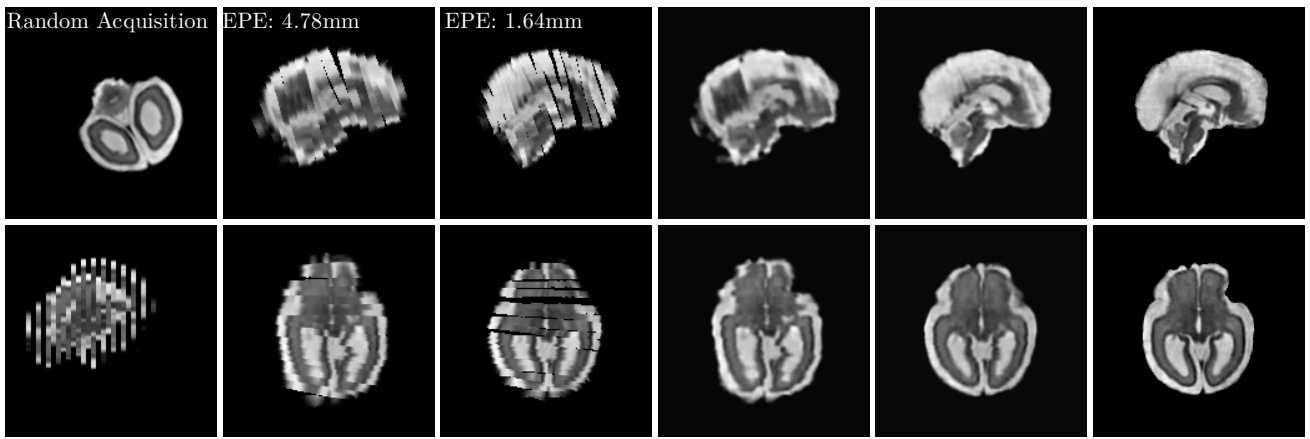

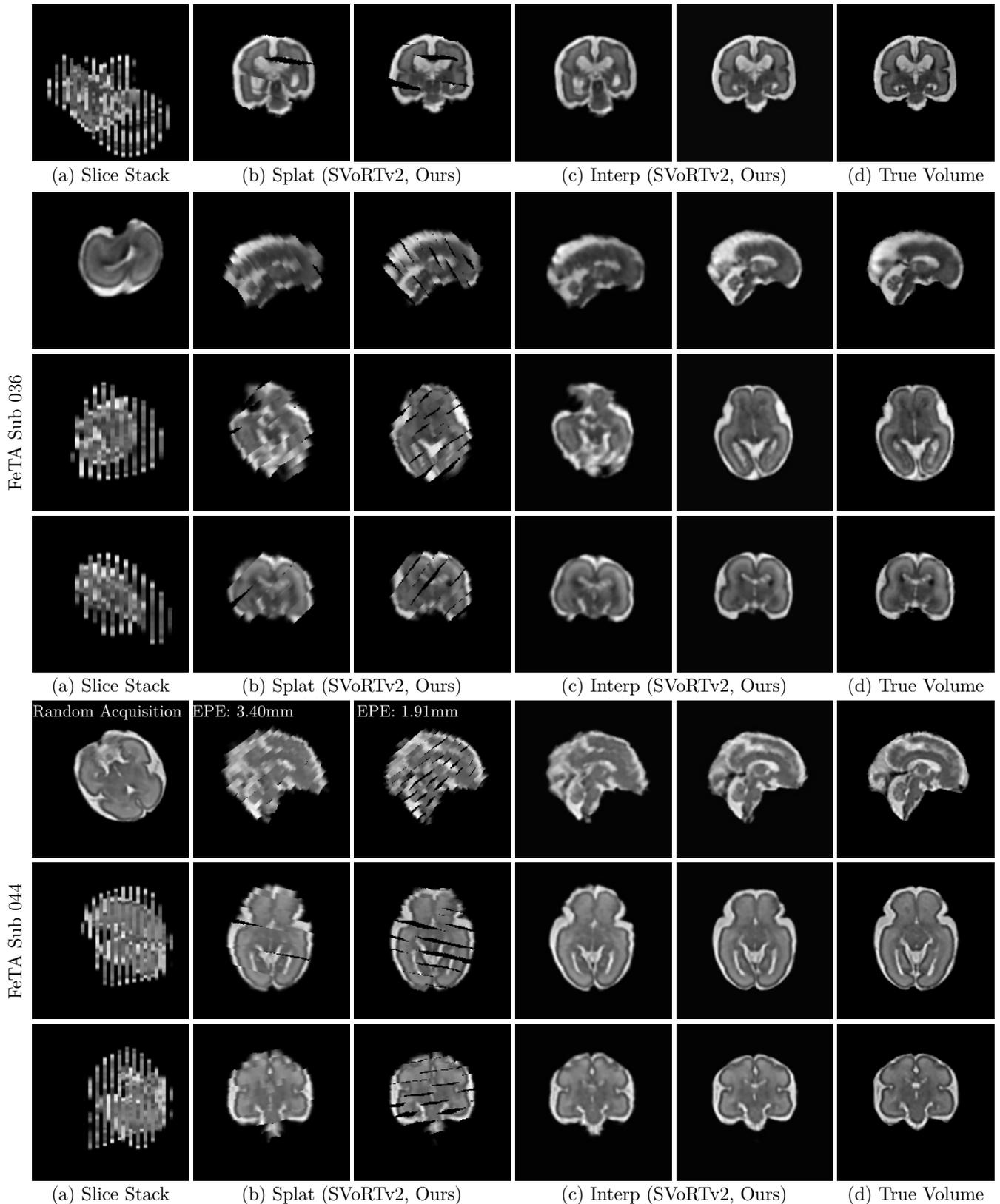

Figure B2: **Single-stack fetal SVR.** We visualize the SVR results on validation subjects from the FeTA dataset [73]. Our results closely resemble the ground truth volumes while SVoRTv2 reconstructions (with our interpolation applied) exhibit spatial distortion from inaccurate slice alignment.

15